\documentclass[sigconf]{acmart}
\renewcommand\footnotetextcopyrightpermission[1]{}
\AtBeginDocument{%
  }

\setcopyright{none}
\copyrightyear{2025}
\acmYear{2025}
\acmDOI{XXXXXXX.XXXXXXX}

\acmConference[ICSE'26]{EEE/ACM International Conference on Software Engineering (ICSE 2026)}{April 12-18 2026}{Rio de Janeiro, BR}
\acmISBN{978-1-4503-XXXX-X/18/06}

\usepackage{placeins}
\usepackage{multirow}
\usepackage{booktabs}
\usepackage{xcolor}
\usepackage{geometry}
\usepackage{color}
\usepackage{dirtree}
\usepackage{booktabs}
\usepackage{graphicx}
\usepackage{makecell}
\usepackage[table]{xcolor} 
\usepackage[justification=centering]{caption}
\usepackage{xspace}
\usepackage{enumitem}
\usepackage{tikz}
\usepackage{dirtree}
\usepackage{tabularx}
\usepackage{ragged2e} %

\newcommand{\rqone}{How are libraries used by their dependent ecosystem?~\xspace}

\newcommand{\rev}[1]{\textcolor{black}{#1}}

\usepackage[strict]{changepage}
\usepackage{framed}

\definecolor{formalshade}{rgb}{0.95,0.95,1}
\definecolor{darkblue}{rgb}{0.0, 0.0, 0.55} %
\newenvironment{quotebox}{%
  \MakeFramed{\advance\hsize-\width\FrameRestore}%
  \noindent\hspace{-4.55pt}%
  \begin{adjustwidth}{}{7pt}%
  \vspace{2pt}\vspace{2pt}%
}
{%
  \vspace{2pt}\end{adjustwidth}\endMakeFramed%
}
\usepackage{hyperref}
\begin{document}

\title{Towards Supporting Open Source Library Maintainers with Community-Based Analytics}

\author{Rachna Raj, Diego Elias Costa}
\affiliation{%
  \institution{REALISE Lab @ Department of Computer Science and Software Engineering\\ Concordia University}
  \streetaddress{P.O. Box 1212}
  \city{Montreal}
  \state{Quebec}
  \country{Canada}
  \postcode{43017-6221}
}
\email{rachna.raj@mail.concordia.ca, diego.costa@concordia.ca}

\begin{abstract}

Open-source software (OSS) is a pillar of modern software development. Its success depends on the dedication of maintainers who work constantly to keep their libraries stable, adapt to changing needs, and support a growing community. Yet, they receive little to no continuous feedback on how the projects that rely on their libraries actually use their APIs. We believe that gaining these insights can help maintainers make better decisions, such as refining testing strategies, understanding the impact of changes, and guiding the evolution of their libraries more effectively.

We propose the use of community-based analytics to analyze how an OSS library is used across its dependent ecosystem. We conduct an empirical study of 10 popular Java libraries and \rev{each with} their respective dependent ecosystem of 50 projects. Our results reveal that while library developers offer a wide range of API methods, only 16\% on average are actively used by their dependent ecosystem. Moreover, only 74\% of the used API methods are partially or fully covered by their library test suite. We propose two metrics to help developers evaluate their test suite according to the APIs used by their community, and we conduct a survey on open-source practitioners to assess the practical value of these insights in guiding maintenance decisions.

\end{abstract}

\maketitle

\section{Introduction}

Modern software systems increasingly depend on open-source software (OSS) \cite{synopsys2024ossra} to deliver feature rich softwares in a cost-efficient and timely manner. Given this widespread presence, maintaining OSS libraries becomes essential to software stability, security and long-term evolution. In this process, library maintainers play a critical role in keeping libraries up-to-date, addressing issues, and adding new features or functionalities to support the broader software systems. 

Maintainers use tools and platforms for dependency management and security, such as Dependabot~\cite{dependabot2024github}, GitHub Notifications~\cite{github2024notifications}, Snyk~\cite{snyk-vulnerability-database}, and Sonatype~\cite{sonatype}. These tools provide alerts, AI-powered risk assessments, and security insights to help manage vulnerabilities and maintain software quality~\cite{9463148}. Beyond dependency alerts, issue tracking platforms~\cite{SnykInvestigatingSecurity} help identify recurring bugs, and automated refactoring tools~\cite{RefactoringMiner} assist in evolving APIs safely.

While these solutions offer valuable support, we believe there is a lack of tools to support library maintainers in understanding how their dependent ecosystem uses their libraries. 
However, there is limited work on analyzing the broader dependent ecosystem to gain deeper insights into library usage. 
Additionally, related work does not explore the applicability of community-based analytics in the maintenance and evolution of library projects. 
Combining API usage analysis with testing practices can offer maintainers valuable feedback, helping them make informed decisions \cite{UnderstandingAPIUsagetoSupportInformedDecisionMakinginSoftwareMaintenance} to prioritize key functionalities, improve API quality, and ensure that key functionalities are well-supported and tested.

To address this gap, we propose leveraging the wealth of data from dependent ecosystems. 
We define the "dependent ecosystem" as a collection of projects that rely on the OSS library as a part of their code to build or run their software.  
To support open-source maintainers, we propose:  
1) analyzing how API methods are used across dependent ecosystem (\textbf{API Usage Analytics});
2) contrasting API usage with library test coverage to show which used methods are tested by their test suite (\textbf{Usage-Based API Test Coverage}); and 
3) assessing the share of dependent projects that rely solely on API methods fully covered by the library test suite (\textbf{Community Test Coverage}).

To evaluate our proposed ideas, we applied community-based analytics on 10 widely popular Java libraries and selected set of 50 dependent projects \rev{per library}. Our analysis of API usage shows that, on average, only 16\% of library public methods are used by the dependent ecosystem.
Among the used API methods, only 7\% are extensively used by the dependent projects, indicating that a small fraction of API methods have significantly higher relevance for dependent projects.  Our analysis of the Usage-Based API Test Coverage shows that, on average, 74\% of used API methods are tested during the libraries' test suite, leaving a sizeable share of 26\% without explicit coverage. Finally, we surveyed open-source practitioners for feedback on the potential usefulness of the analytics and proposed metrics. 
Feedback from 20 respondents showed significant interest in tools that track dependent API usage and their contrast with the test coverage insights. While opinions on community test coverage (CTC) were mixed, participants responded positively to the idea of testing plan recommendations and offered valuable suggestions for improvement.

Our paper contributes to the extensive body of work in open-source software development.
This paper makes the following key contributions: 
\begin{itemize}[noitemsep, topsep=0pt, leftmargin=*]
    \item We propose a series of community-based analytics to support open-source library maintainers. The analytics include API usage frequency distribution and two novel test coverage-related metrics: 
    \begin{itemize}[noitemsep, topsep=0pt]
        \item \emph{Usage-Based API Test Coverage:} Reports the share of API methods used by the dependent ecosystem that is covered partially or fully by the library test suite. 
        \item \emph{Community Test Coverage:} Reports the share of dependent projects that rely solely on API methods fully covered by the library test suite. 
    \end{itemize}
    \item We conduct an empirical study on 10 popular Java libraries and their top 50 dependents \rev{each} to evaluate these analytics.
    \item We surveyed open-source practitioners on the usefulness of our proposed analytics and metrics.
    \item We made all data collected and scripts developed available in our replication package\cite{HelpingLibraryMaintainers}. 
\end{itemize}

\section{Related Works}
\rev{Previous studies have explored 1) how dependents use APIs of software libraries, 2) the testing practices of open source software libraries and 3) the potential for developing community-based metrics for open source software development.}

 \rev{\textbf{API usage in OSS:} A substantial body of research has examined how APIs are used in open-source software (OSS), shedding light on how library features are adopted by their dependent ecosystems \cite{bauer2014exploratory, meng2019developers, de2013multi}. These efforts often aim to improve API documentation, inform design decisions, or enhance developer productivity \cite{de2013multi}.}

Bauer et al. \cite{bauer2014exploratory} conducted an exploratory study of API reuse in large industrial systems, while Nguyen et al. \cite{nguyen2016learning} applied a statistical model over bytecode to recommend completions for incomplete API usage. Qiu et al. \cite{qiu2016understanding} analyzed 5,000 OSS Java projects and found that many API methods, classes, and fields are never used—an observation consistent with our findings. Similarly, Sawant et al. \cite{sawant2017fine} proposed fine-GRAPE to mine API usage in five popular Java libraries and showed that a small set of core features dominate usage, often introduced in early API versions.

These studies consistently suggest that libraries tend to be adopted for a narrow subset of their capabilities, with many features remaining underused.
Harrand et al. \cite{harrand2022api} extended prior work to the ecosystem level, analyzing API usage across popular Maven libraries. From the client side, they found that 41.12\% of declared dependencies were unused at the bytecode level, exposing a significant disconnect between declared and actual usage. From the library side, they tracked how public APIs are used across versions and showed that many methods are rarely needed in practice. Their findings support API streamlining as a viable strategy to reduce maintenance overhead without disrupting client needs.

\rev{While these works offer valuable insights for guiding migration and maintenance, they do not explore fine-grained, method-level analytics that identify which API elements are most relied upon across dependent projects. Such insights are essential for informing API evolution decisions. Furthermore, actionable usage analytics grounded in real-world ecosystems have yet to be effectively presented to maintainers, leaving the potential for optimizing library maintenance and development efforts in a data-driven manner.}

\textbf{Testing practices and coverage in OSS:} The challenges of testing in open-source software (OSS) have been widely studied. Lu et al.\cite{lu2020research} argue that OSS projects, often developed by diverse contributors, lack consistent testing strategies due to the fast pace of development. A survey by Zhao and Elbaum\cite{zhao2003quality} found that 80\% of OSS developers had no formal testing plans. Kochhar et al.\cite{pavneet2013empirical} analyzed over 20,000 GitHub projects, showing that 84.87\% had fewer than 100 test cases, and only 2.57\% had more than 500. Takasawa and Sakamoto\cite{takasawa2014open} introduced test measurement indexes through an analysis of over 750 Maven projects, revealing that just 14.8\% passed all their tests. They emphasized \textbf{the need for practical test quality metrics}, such as pass/fail rates or lightweight coverage measures that do not require test execution.

These findings motivate our focus on \textbf{community-based coverage metrics} in OSS library projects. Although code coverage is widely used to assess test adequacy and is often treated as a proxy for quality assurance~\cite{khatami2023stateofthepracticequalityassurancejavabased}, its limitations are well documented. Kochhar et al.\cite{kochhar2014empirical} found only a weak correlation between coverage and post-release defects. Inozemtseva et al.\cite{inozemtseva2014coverage} similarly reported that coverage is not strongly tied to test effectiveness. Yet they note that it remains useful for revealing under-tested parts of the codebase.

\rev{A recent study by Zaki et al.~\cite{zaki2025understanding}, which closely aligns with our work, examines the gap between API usage and testing in C libraries. Their findings resonate with ours and further underscore the need for coverage metrics that reflect real-world usage.}

\rev{\textbf{Community Metrics and tools for OSS maintainers: } A range of tools supports open-source maintainers in managing development and ensuring software quality. Services like Dependabot~\cite{dependabot2024github}, GitHub Notifications~\cite{github2024notifications}, Snyk~\cite{snyk-vulnerability-database}, and Sonatype~\cite{sonatype} provide alerts, AI-assisted risk assessments, and security insights to help address vulnerabilities and maintain code health~\cite{9463148}. Issue tracking platforms~\cite{SnykInvestigatingSecurity} help identify recurring bugs, while automated refactoring tools such as RefactoringMiner~\cite{RefactoringMiner} assist in evolving APIs safely.}

\rev{Beyond tooling support, prior work has examined OSS communities in terms of contributor roles, participation trends, and engagement patterns~\cite{10.1109/TSE.2014.2349496}, offering valuable insights into who contributes and how projects evolve. However, these studies often overlook the broader community of code consumers. While much research focuses on contributors and code maintainability, it rarely considers how APIs are actually used by dependent projects. A survey study by Wattanakriengkrai et al \cite{10685609} suggests that maintainers need tools or metrics to support OSS and sustain these ecosystems.}

\rev{We attempt to address this gap with two community-based metrics: Community Test Coverage and Usage-Based Test Coverage. These metrics highlight what the community actively consumes and how this insight can support maintainers.}\rev{ Building on prior research, we contrast API usage with test coverage and propose usage-informed, community-level metrics to support OSS maintenance and evolution.}

\begin{figure*}
    \centering
    \includegraphics[width=.8\textwidth]{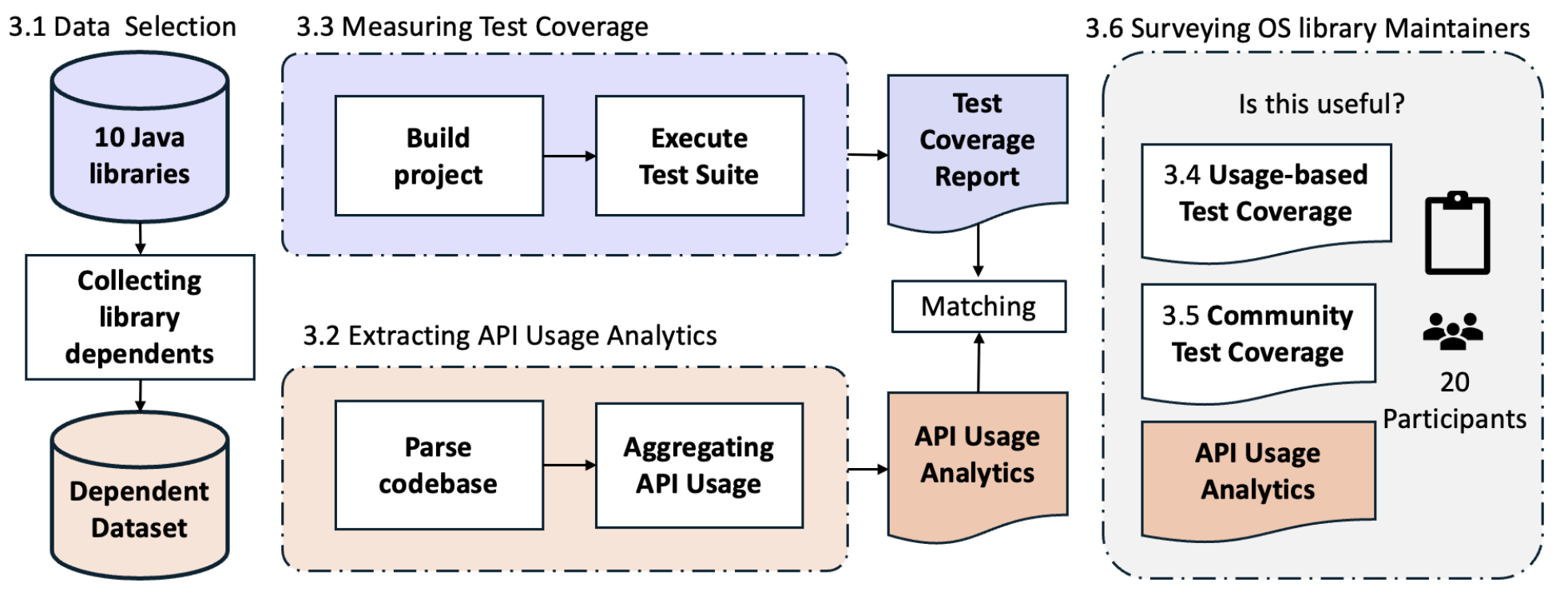}
    \caption{Overview Diagram of this project}
    \label{fig:Overview}
\end{figure*}

\section{Methodology}

The goal of our study is to extract community-based analytics from 10 popular library projects and evaluate their usefulness for maintenance and evolution.
Following the approach in Figure~\ref{fig:Overview}, we selected 10 popular Java OSS libraries and, for each, 50 dependent projects (Section~\ref{DataSelection}).
We mined API usage data (Section~\ref{usagepattern}) and extracted test coverage from the libraries' test suites (Section~\ref{testcoverage}).
Combining these, we defined two community-based metrics: Usage-Based Test Coverage (Section~\ref{sub:usage-based-api-test-coverage}) and Community Test Coverage (Section~\ref{sub:community-test-coverage}).
Finally, we surveyed OSS maintainers (Section~\ref{survey}) to assess the perceived usefulness of these analytics for guiding project maintenance and evolution.

\rev{Our study design and methodology are designed to answer the following research questions:}

\rev{\textbf{API Usage Analytics (Section 4)}}
\begin{itemize}[noitemsep, topsep=0pt, ]
    \item \rev{How are libraries used by their dependent ecosystem?}
    \item \rev{Is API Usage Analytics useful for maintainers?}
\end{itemize}

\rev{\textbf{Usage-based API Test Coverage (Section 5)}}
\begin{itemize}[noitemsep, topsep=0pt]
    \item \rev{How often are used API methods covered by tests?}
    \item \rev{Is Usage-Based API Test Coverage useful for maintainers?}
\end{itemize}

\rev{\textbf{Community Test Coverage (Section 6)}}
\begin{itemize}[noitemsep, topsep=0pt]
    \item \rev{How often are dependent projects using only APIs fully covered by tests?}
    \item \rev{Is Community Test Coverage useful for maintainers?}
\end{itemize}

\subsection{\textbf{Data Selection}}  
\label{DataSelection}

\textbf{Library project selection:}
\rev{The goal of our study is to evaluate the usefulness of the proposed metrics for maintainers of widely-used libraries. 
We focus on Java libraries because it is widely adopted \cite {tiobe2024index}, and has a statically typed nature that allows us to resolve variable types before compilation, which is crucial for accurately distinguishing the API usages.
To create a high-quality candidate pool, we began selecting the top most depended upon 150 Java libraries on Libraries.io. We then applied the filtering steps:
}

\begin{itemize}[noitemsep, topsep=0pt, leftmargin=*]
    \item \rev{\textbf{Hosted on GitHub Platform.} We remove 10 library projects hosted outside of GitHub, as we base further steps of our methodology (dependent analysis) on the GitHub platform.} 
    \item \rev{\textbf{Popular and mature library projects.} We filtered out libraries with $\leq$ 40 contributors and $\leq$ 10 years of age to focus on libraries with long history of collaborative development. This reduced our pool to 103 libraries.}
    \item \rev{\textbf{Excluding frameworks and Development Tools:} 
    Infrastructure heavy libraries (e.g., frameworks like Spring Boot) or build tools would require complex handling of annotations, dependency injection, and reflection aspects known to challenge static analysis tools \cite{7985689, laigner2022cataloging, sun2022analyzingimpactdependencyinjection}. 
    This reduced our pool to 64 libraries.}
    \item \rev{\textbf{Github "used by" metadata:} We excluded 21 libraries that did not expose the "Used by" information in their GitHub repository, as we use this information to find dependent projects (e.g.Lombok, Joda-Time), resulting in 43 candidates.}
\end{itemize}

\

\rev{After this filtration, we are left with 43 candidates that fulfill the technical requirement and are popular and well-established software projects. 
We group the selected libraries based on their Maven category~\cite{di2023hybridrec}, a metadata provided Maven repository website to inform the community about the library domain.
We then employ a \textbf{purposive sampling} \cite{baltes2021samplingsoftwareengineeringresearch} to select 10 categories, and one library sample for each category.
Since we analyze test coverage, we require each library to build successfully in a controlled environment. 
Four libraries failed to build in our environment (e.g, Gson) and were replaced by another library from the same category (e.g., Jackson-databind). 
Two database libraries (e.g., HikariCP, H2 Database) were removed from our study due to failing to build due to unmet dependencies. 
Hence,based on the above criteria relevant to our study, we selected 10 libraries similar to prior work that analyzed a smaller set of projects~\cite{inozemtseva2014coverage}.
Table \ref{tab:library-selection} lists the selected libraries with their quantitative community stats.}

\textbf{Collecting library dependents:} 
\rev{GitHub’s “Used by” section includes many toy or inactive repositories, not representative of the software project dependents we aim to analyze~\cite{Kalliamvakou:GitHub}. To use dependents as meaningful feedback signals for library maintainers, we focused on selecting active, real-world dependent projects and applied the following filtering criteria:}

\rev{\textbf{Popularity and maturity:} We included only projects with $\geq$ 100 GitHub stars, >3 contributors, and $\geq$ 5 years of age, aligning with prior work to reflect stability and community interest~\cite{han2019characterization}.}

\rev{\textbf{Timeliness (version alignment):} 
Most dependents tend to depend on outdated versions (technical lag~\cite{kula2018developers,jafari2023dependency,derr2017keep}), which would lead to noise in the community signal (e.g., usage of deprecated APIs).
To ensure that our findings are relevant to ongoing development and maintenance efforts by the library maintainers, we included only dependents using the latest major version stream of the library (e.g., depends on Awaitility version 4.2.x), as declared in their \texttt{pom.xml}, to ensure our findings reflect current usage.
}

\begin{figure} 
    \centering
    \includegraphics[ width=1.0\linewidth]{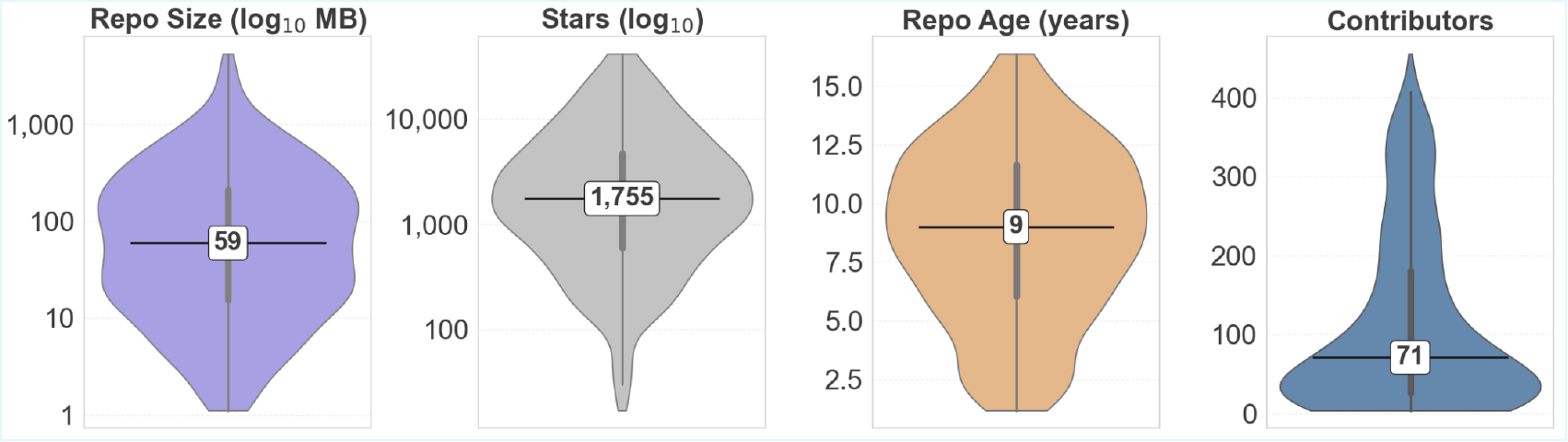}
    \caption{\rev{Distribution of dependent's repository metrics across a representative datasets for 10 candidate libraries.}}
    \label{fig:dependents_distribution}
\end{figure}

\rev{Applying these filters, especially version alignment, significantly reduced the number of eligible dependents for some libraries (e.g., Httpcomponents-client, Logback, JUnit4, Guava), as most of their dependents had technical lag. To maintain comparability, we capped the number of selected dependents per library at 50.}
\rev{Figure \ref{fig:dependents_distribution} represents the quantitative data for the selected dependents sample across 10 libraries}

\begin{table*}
    \caption{\rev{The selected 10 Java Libraries projects, their project statistics and dependent exemplaries.}}
    \label{tab:library-selection}
    \centering
    \begin{tabular}{@{}p{3.1cm} p{2.9cm} p{0.6cm} p{1cm}|r r r|r@{}}
    \toprule
        &\rev{\textbf{Maven}} & \rev{\textbf{Maven}} & & \multicolumn{3}{c}{\rev{\textbf{Library Project Statistics}}} & \\
        \rev{\textbf{Library}} & \rev{\textbf{Category}} & \rev{\textbf{Rank}} & \rev{\textbf{Version}} & \rev{\textbf{Contrib.}} & \rev{\textbf{Stars}} & \rev{\textbf{Age (Y)}} & \rev{\textbf{Example Dependents}} \\
    \midrule
        \rev{AssertJ} & \rev{Assertion} & \rev{1} & \rev{3.26.x} & \rev{360} & \rev{2.6k} & \rev{11} & \rev{Hadoop, Javaparser} \\ 
        \rev{Awaitility} & \rev{Concurrency} & \rev{3} & \rev{4.2.x} & \rev{44} & \rev{3.8k} & \rev{15} & \rev{Spring-cloud-gcp, Geoserver} \\ 
        \rev{Commons Codec} & \rev{Encoding utilities} & \rev{1} & \rev{1.17.x} & \rev{47} & \rev{457} & \rev{15} & \rev{Feldera, Kylin} \\
        \rev{Google Guava} & \rev{Core utilities} & \rev{1} & \rev{33.3.x} & \rev{309} & \rev{50k} & \rev{10} & \rev{Jenkins, Neo4j} \\ 
        \rev{HttpComponents Client} & \rev{Http clients} & \rev{1} & \rev{5.4.x} & \rev{92} & \rev{1.5k} & \rev{15} & \rev{Keycloak, OpenFeign} \\
        \rev{Jackson Databind} & \rev{JSON Processing} & \rev{1} & \rev{2.18.x} & \rev{247} & \rev{3.5k} & \rev{13} & \rev{Swagger-inflector, Qbicc} \\
        \rev{Javassist} & \rev{Byte-code Engineering} & \rev{1} & \rev{3.20.x} & \rev{43} & \rev{4.1k} & \rev{11} & \rev{HikariCP, Dropwizard} \\
        \rev{Jsoup} & \rev{Parsing Libraries} & \rev{1} & \rev{1.18.x} & \rev{108} & \rev{10.9k} & \rev{15} & \rev{Quarkus, Hibernate-search} \\
        \rev{JUnit4} & \rev{Unit Testing} & \rev{1} & \rev{4.13.x} & \rev{154} & \rev{8.5k} & \rev{16} & \rev{SpaceVim, Akka} \\  
        \rev{Logback} & \rev{Logging} & \rev{2} & \rev{1.5.x} & \rev{123} & \rev{3k} & \rev{15} & \rev{Helios, Eclipse-store} \\  
    \bottomrule
    \end{tabular}
\end{table*}

\subsection{\textbf{Extracting API Usage Analytics}}
\label{usagepattern}

We want to analyze the contrast between all publicly available API methods exposed by library projects and the API methods that are used by the dependent ecosystem.

\noindent
\textbf{Extracting all publicly available API methods:}
We determine the set of \textit{publicly available APIs} by parsing the library bytecode from the JAR files. 
We use the Javap tool \cite{javapdoc} provided by the JDK to disassemble bytecode and traverse the content of a JAR file. 
To filter out only public methods, we first use the \texttt{jar -tf} command to list all .class files in the JAR. 
Then, for each class, we process the output using the \texttt{javap -public} flag to collect the package name, class name, and public method signatures, storing them in a structured format.

\noindent
\textbf{Extracting the used API methods:}
To extract the used API methods for each library, we clone the latest snapshot of each of their dependent projects. 
Given that we have to analyze a total of 500 dependent projects (50 dependents per library project), we opt to parse their code using a two-step approach: 

\begin{enumerate}[noitemsep, topsep=0pt, leftmargin=*]
    \item \textbf{Analyse file dependencies:} To reduce the number of files to be parsed, we first analyze only the import statements of all Java files.
    We skip the file from the second parsing step if no import statement refers to the target library. 
    \item \textbf{Extract all target library method calls:} In this step, we extract all relevant API method calls from the dependent projects. 
    We use JavaParser \cite{javaparser}, which is a mature parser with support for recent Java versions and has been used in similar studies in software engineering \cite{armbruster2022parsing, dhalla2020performance,hosseini2013javaparser} to obtain the abstract syntax tree (AST) of \rev{the} Java codebase. We traverse the AST to extract the API references of library method calls.
    As methods can share the same class and name (method overloading), we must resolve variable types to properly identify a method call.
     
    To resolve the types during the parsing process, we use JavaParser's JarTypeSolver \cite{JarTypeSolver}, which requires JAR files to accurately resolve method calls and variable types, particularly concerning third-party library types. 
    We provide the JAR file needed by the type-solver during the parsing process.
\end{enumerate}

At the end of this process, our parser yields a list of method calls of the target library for a single dependent project, including relevant metadata, such as the fully qualified method name, file location, and method parameters.

\textbf{Aggregating API usage analytics:}  
We define \textit{API usage analytics} of a library as the collection of all API method calls to the library extracted from the dependent projects. The API usage \( U_L \) of a library \( L \) \rev{is shown} as:

\[
U_L = \biguplus_{i=1}^{n} M_{D_i}
\]

where \( M_{D_i} \) represents a multiset (collection with potential duplicates) of method calls in dependent project \( D_i \). Each call is identified by its fully qualified method name, package, and class. The collection \( U_L \) aggregates all method calls across the 50 dependent projects.
At the end, we have a multiset union of all method calls extracted from the dependent projects, with their respective metadata, which allows us to understand the distribution of the frequency of API usage.

\subsection{Measuring Test Coverage}
\label{testcoverage}

One area of software maintenance where API Usage analytics can be insightful is in guiding testing practices. 
We are interested in contrasting the API Usage with the test coverage of each library project, to evaluate how often library test suites cover used API methods.  
Measuring test coverage can be a challenging task due to the problems related to building a software project, such as compilation errors \cite{khatami2023stateofthepracticequalityassurancejavabased} or build processes sensitive to environmental factors~\cite{Horvath2019}.
To account for these challenges, we proceed to measure the test coverage following a multi-step process:

\noindent
\textbf{Building the project}: We begin by reviewing each library’s project documentation to determine the correct build configuration. Once the projects were successfully built, we identify the configuration needed to run the full test suite. While we were able to set up most libraries, we faced some challenges, such as missing modules referenced, environment-specific requirements, which we resolved to ensure proper execution. For libraries with larger test suites, like Guava's, we had to adjust the heap space allocation due to the large number of tests. After configuring the heap space correctly, all test suites ran smoothly. All library projects rely on Maven to run the project test suite.

\textbf{Executing libraries test suite:}
We used the open source code coverage tool Jacoco \cite{hoffmann2009jacoco} because it is widely used, actively maintained, and has been adopted in previous research \cite{hoffmann2009jacoco, silva2023flacocofaultlocalizationjava, Horvath2019}. It works by instrumenting Java bytecode, supports multiple Java versions and integrates easily with Maven, making it a practical choice for our study.

We specifically focus on extracting statement coverage, which indicates which program instructions were executed during tests and which were not. Since JaCoCo operates at the bytecode level, our coverage data reflects the number of bytecode instructions exercised by the tests, which closely aligns with source code line execution. Statement coverage is commonly used because it is easy to interpret \cite{inozemtseva2014coverage} and ensures that if a method is marked as covered, at least one of its instructions was executed during testing \cite{jacoco_counters}.

For multi-module projects like Guava, we ran JaCoCo on each module separately to collect coverage data for the entire library. This approach allowed us to gather detailed coverage data for each module, which was then combined to provide overall coverage for the entire library.
\rev{To minimize the impact of flaky coverage results, we executed the test suite for each library three times, as suggested by prior works~\cite{soto2023coverage, Horvath2019} to obtain reliable coverage data. For our selected libraries, we got consistent coverage results across the three runs, indicating minimal influence of random test failures or inconsistencies.}

\textbf{Matching test coverage data to API usage: } 
After collecting the coverage data, we need to map the test coverage data with the API Usage analysis we extracted for each library.
We cross-reference the API usage analytics metadata with the test coverage report.
The API usage analytics contain the full details of API method calls (class, fully qualified name, parameter types); however, the coverage report may omit the parameter types in some scenarios, e.g., generic methods, parameter polymorphism, or methods that accept interface types.
To account for these cases, we implemented a hierarchical matching algorithm:

\begin{enumerate}[noitemsep, topsep=0pt, leftmargin=*]
\item \textbf{Full match}: The method class, fully qualified name (FQN), and parameters match the coverage reports.
\item \textbf{Partial match with no ambiguity}: The method class, fully qualified name matches, and the number of parameters in both reports are consistent. While the parameter type does not match (e.g., String and Object), there is no ambiguity in the matching because there is no other method in the coverage report with the same number of parameters.  
\item \textbf{Partial match with ambiguity}: The method name and the number of parameters match, but there is more than one method possible for the matching due to method overloading. In this case, static analysis cannot fully match the methods, and we resort to mapping the methods to the highest coverage value found (\rev{upper}-bound coverage estimation).
\item \textbf{No match}: If neither the method name nor the number of parameters match, the method is not found in the coverage report.
\end{enumerate}

In our analysis, we find that 84\% of the API methods fall into cases 1 and 2, where there is no ambiguity in matching. 
Partial matching with ambiguity represented 14\% of the cases, and 2\% of the methods from the API Usage Analytics did not find a matching in the test coverage.
Upon manual inspection, we find that in these cases, the coverage simply did not include an entry for the method, and we remove these cases from the test coverage analyses.

\subsection{Usage-Based API Test Coverage}
\label{sub:usage-based-api-test-coverage}

\noindent
\textbf{Motivation and core idea.} 
Traditional test coverage metrics, such as line or branch coverage, may overlook the extent to which a library's API methods are exercised.
We propose a complementary test coverage metric that helps library maintainers identify gaps in their testing suite by reporting API methods that 1) are used by their dependent ecosystem and 2) are not explicitly tested during the test suite execution.

\textbf{Computing usage-based API test coverage:}
The usage-based API test coverage is defined by the share of API methods that are used by the dependent ecosystem and \rev{are} partially or fully covered by the library project test suite. 
It can be expressed as a percentage calculated using the formula:

\begin{equation}
\text{Usage-Based Cov (\%)} = \left( \frac{N_{\text{Covered}}}{N_{\text{used}}} \right) \times 100
\label{eq:ubc}
\end{equation}

\noindent where:
\begin{itemize}[noitemsep, topsep=0pt]
    \item \( N_{\text{Covered}} \): The number of API methods that are fully or partially covered by any test case.
    \item \( N_{\text{used}} \): The number of used API methods found in 50 dependent projects.
\end{itemize}

\subsection{Community Test Coverage}
\label{sub:community-test-coverage}

\noindent
\textbf{Motivation and core idea:} 
While traditional test coverage metrics focus on the library's internal tests, they may fail to capture the broader impact on dependent projects.
We propose a metric to assess the extent to which dependent projects rely solely on API methods that are minimally tested by the library’s test suite.

\noindent
\textbf{Computing community test coverage: }
This metric evaluates the extent to which the test suite of a library fully covers all the API methods used by its dependent projects.
CTC is computed as the percentage of dependent projects where all the API methods they use are fully covered by tests in the library. Formally, it is expressed as:

\begin{equation}
\text{CTC (\%)} = \left( \frac{NP_{\text{FullyCovered}}}{NP_{\text{total}}} \right) \times 100
\label{eq:CTC}
\end{equation}

\noindent where:
\begin{itemize}[noitemsep, topsep=0pt]
    \item \( NP_{\text{FullyCovered}} \): The number of dependent projects where the library test suite fully covers all the API methods they use. 
    \item \( NP_{\text{total}} \): The total number of dependent projects analyzed.
\end{itemize}

To compute this, we first grouped the data obtained from matching API usage with coverage, based on dependent projects, to identify the API methods used by each project and their corresponding coverage values. During this step, we excluded dependent projects without a valid coverage score, which slightly reduced the number of dependents analyzed for some libraries. Next, we evaluated each dependent project to count the fully covered API methods. A dependent project was considered "fully covered" if all the API methods it used had 100\% coverage. While capturing coverage, we focus on the API method without considering the calling chain to gain a simple and straightforward coverage assessment.

\subsection{Surveying Open Source Library Maintainers}
\label{survey}

\rev{We conducted a survey to collect feedback from open-source practitioners (maintainers and contributers).}
In particular, we analyse their perspective on the levels of usefulness of the proposed community-based analytics in supporting their library project's maintenance and evolution.

\begin{table}
    \caption{Developer and OSS Contribution Experience}
    \vspace{-.2cm}
    \label{DeveloperExp}
    \centering
    \begin{tabular}{|lcc|lcc|}
    \hline
    \textbf{Developer Experience} & \multicolumn{2}{c|}{\textbf{\#}} & \textbf{OSS Experience} & \multicolumn{2}{c|}{\textbf{\#}} \\
    \hline
    Less than a year & 0 &  & Less than a year & 1 &  \\
    Between 1 to 5 years & 4 &  & Between 1 to 3 years & 1 &  \\
    Between 5 to 10 years & 2 &  & Between 3 to 5 years & 2 &  \\
    More than 10 years & 14 &  & More than 5 years & 16 &  \\
    \hline
    \end{tabular}
    \vspace{-.2cm}
    
\end{table}

\textbf{Survey design:} To obtain feedback from broader audience, we designed two types of surveys: 1) A specialized survey tailored to the specific ten libraries we analyzed, including the real analytics extracted from the empirical study, to help maintainers contextualize their usefulness to their library project. 2) A general survey, designed for broader audience of OSS contributors and maintainers. Here, we included hypothetical examples based on our study findings to make the survey accessible to contributors from various open-source projects.

Before participants answer any questions, we explained our approach to collecting community-level statistics and presented scenarios to asked three to four agree/disagree questions for each. Figure \ref{fig:SurveyResults} presents all questions that were part of our survey. \rev{We focused on three key areas:}

\begin{itemize}
    \item \textbf{API usage analytics:} How API usage analytics influence maintainers' decisions on updates, bug fixes, and testing.
    \item \textbf{Usage-based test coverage:} How maintainers perceive usage-based test coverage metrics to improve maintenance activities.
    \item \textbf{Community test coverage \& testing plan recommendations:} How maintainers perceive about Community test coverage metric and suggested recommendations to improve maintenance activities.
\end{itemize} 

\rev{These areas evaluated maintainers' opinions on the usefulness of the proposed analytics for understanding API usage and improving test coverage. 
Here we also ask participants whether they conduct similar analysis in their library projects (Q1-2), followed by two to three agree/disagree questions per area on a linear scale (Q3-15). 
The purpose of these questions is to help us understand whether maintainers find community-based insights effective for their maintenance tasks. In all these questions, we asked participants to rate their level of disagreement/agreement on a 5-point Likert scale (strongly disagree, disagree, neutral, agree, strongly agree). Additionally, we incorporated four optional open-ended questions in each section to gather more insights from maintainers' perspectives.}

\rev{The survey also included two background questions about participants' software development and open-source contribution experience, as developers may have significantly more experience with software development in general than with OSS contribution.
}
To ensure ethical transparency, we informed participants about their right to withdraw their responses at any time before the survey results were published in research venues. Participants had to provide explicit consent before proceeding, and they also had the option to submit the form without answering all the questions.

\textbf{Participant recruitment:} \rev{We recruit potential participants via each library’s GitHub contributor dashboard, prioritizing active contributors to ensure recent familiarity. For the specialized survey,
we select one active maintainer and one top contributor (from 2024 or later) per project. 
To leverage the availability of multiple active contributors in some cases, we extended the outreach to 3 candidates in five libraries, namely (Apache Commons Codec, HttpComponents Client, Guava, JUnit, Jackson Databind), resulting in 25 personalized invitations.}
\rev{In total, we invited 25 participants for the specialised survey on 10 libraries and received five responses, achieving a \textbf{20\% response rate}}.
    
\rev{For the general survey, we contacted contributors and/or maintainers of popular open-source Java library projects hosted on GitHub. Some examples are Spring, Netty, Kotlin, Byte Buddy, and Gradle. A full list of libraries included in the general survey is present in our replication package~\cite{HelpingLibraryMaintainers}. 
In total, we sent emails to 51 maintainers/contributors using a general survey and received 15 responses, resulting in a \textbf{30\% response rate}}.
As seen in Table~\ref{DeveloperExp}, all participants have more than 2 years of developer experience, with 14/20 participants having 10+ years of experience. 
Respondents also have a reasonably long history of maintaining OSS, with 16 participants reporting more than 5 years of experience.

\rev{\textbf{Analysis of survey results:} Figure \ref{fig:SurveyResults} presents our survey questions and their results. 
The figure shows different questionnaire categories, statements, and distribution of agreement levels. 
At the top of the chart (Q1 and Q2), we present a stacked bar chart that represents two single-select questions; we show the number of responses for each option highlighted in different shades of green.
We used a Likert scale to analyze all agreement/disagreement questions, and present the results as a divergent stacked bar, with the stacked bar centred on "Neutral" (grey). 
Hence, the statements with mostly blue tending to the right show agreement among the participants. Conversely, statements with a bar plot tending to the left (red) convey more frequent disagreement.}

\rev{
For open-ended responses, we applied thematic analysis. Two coders independently analyzed a total of 35 responses from four open-ended questions. 
To measure inter-rater agreement, we used Krippendorff’s alpha\cite{ krippendorff2011computing}, as it properly handles small sample sizes and allows for multiple code categories~\cite{marzi2024k}.
The two coders obtained an agreement score of 0.82 (strong agreement). 
After the initial coding, the coders met to discuss and resolve any disagreements. 
In Table \ref{tab:tikz-thematic-summary}, we provide the themes decided for open-ended responses in our survey.}

\section{API Usage Analytics}

\subsection{\rqone}
\label{sub:api-analytics}

The goal of this question is to report how dependent projects use the API methods of each library.
We report 1) the share of used API methods and 2) their distribution across the dependent projects.

\begin{figure}
    \centering
    \includegraphics[ width=1.0\linewidth]{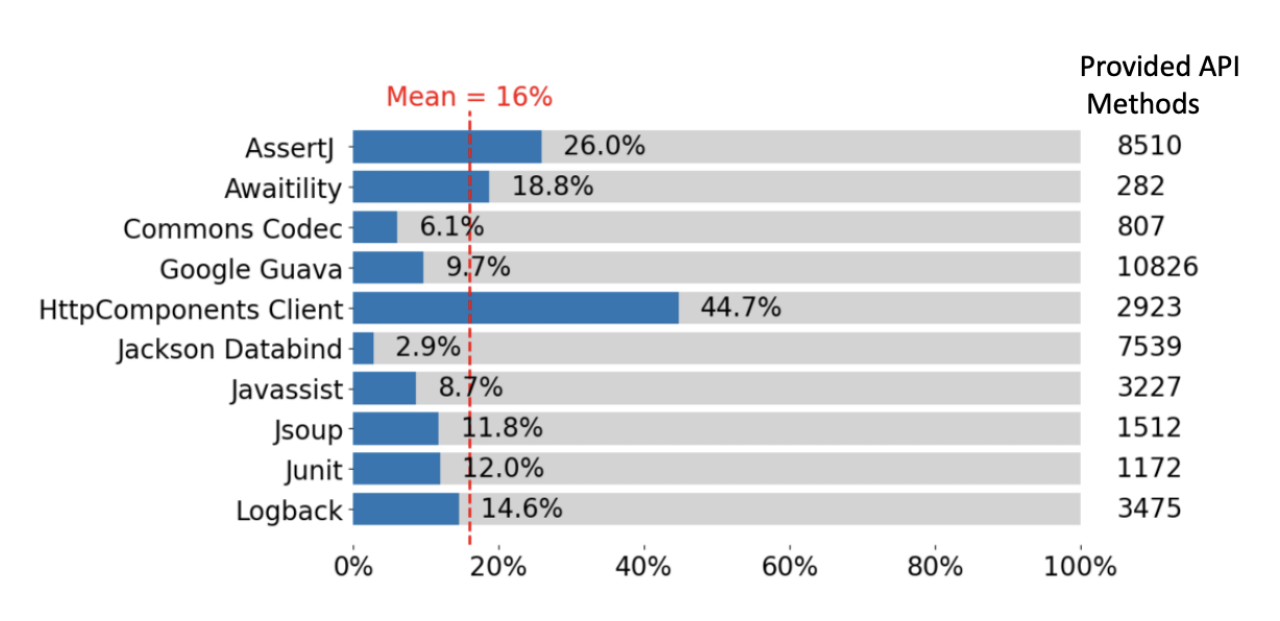}
    \vspace{-.2cm}
    \caption{Percentage of the total available API methods used by the 50 dependent projects.
    We highlight the mean percentage across libraries with a vertical dotted mark.}
    \label{fig:rq1_distribution}
\end{figure}

\textbf{How many API methods are used?} In Figure \ref{fig:rq1_distribution}, we present the percentage of used APIs across the 10 evaluated library projects. 
While libraries often expose thousands of public API methods, on average, only 16\% of API methods are used by the library's dependents. 
Some libraries show significantly higher usage. For example, dependents of Apache HttpComponents Client use approximately 1,308 out of the 2,923 available methods, covering nearly 45\% of the total API. Similarly, AssertJ-core has a usage rate of 26\%, with 2,210 unique API methods utilized.

In contrast, libraries like jackson-databind and commons-codec offer more API methods, but analysis of 50 dependents shows lower usage of unique API methods. This suggests that a higher usage percentage indicates a broader range of features being used, whereas a lower percentage reflects more specialized or focused usage.

\begin{table}
\centering
\caption{Usage distribution of selected library projects based on their top 50 dependent clients}
\vspace{-.2cm}
\begin{tabular}{@{}l|r|rrrr}
\toprule
        &    \textbf{\# used}            & \multicolumn{4}{c}{\textbf{Used by \# dependent proj.}} \\
\textbf{Library}  & \textbf{APIs} & 1  & 2-4  & 5-9  & 10+  \\
\midrule
AssertJ & 2210 & 77.6\% & 14.2\% & 4.2\% & 4.5\% \\
Awaitility & 53 & 28.3\% & 34.0\% & 15.0\% & 26.4\% \\
Commons Codec & 49 & 53.1\% & 28.6\% & 10.2\% & 6.1\% \\
Google Guava & 1046 & 51.4\% & 37.5\% & 7.5\% & 5.0\% \\
HttpComponents Client & 1308 & 82.9\% & 13.3\% & 2.4\% & 1.6\% \\
Jackson Databind & 216 & 57.8\% & 31.0\% & 7.4\% & 3.7\% \\
Javassist & 281 & 49.5\% & 37.0\% & 5.3\% & 8.5\% \\
Jsoup & 179 & 42.5\% & 39.7\% & 11.7\% & 7.2\% \\
JUnit4 & 103 & 43.7\% & 39.8\% & 8.7\% & 10.6\% \\
Logback & 509 & 59.6\% & 32.0\% & 5.6\% & 3.1\% \\

\midrule
\textbf{Mean} & \textbf{595.4} & \textbf{54.6\% }& \textbf{30.8\%} & \textbf{7.8\%} & \textbf{7.3\%} \\
\bottomrule
\end{tabular}

\label{tab:RQ1_heuristic}
\vspace{-.2cm}
\end{table}

\textbf{What is the API usage distribution on dependent projects?}
Table \ref{tab:RQ1_heuristic} presents the API usage information for each library. 
Our initial observation is that most API methods across the 10 libraries fall into the low-usage category, meaning they were used in only one project. For most libraries, at least 42\% of their API methods are in this low-usage group, indicating limited adoption across projects. Notably, the Apache HttpComponent Client library has 82\% of its API methods (1,085 out of 1,308) used by only one dependent project. Similarly, AssertJ has 77\% of its API methods (1,717 out of 2,210) in this category, reinforcing the trend of limited usage across dependents.  This is particularly interesting because these libraries offer a broad range of API methods, but most still see limited usage, primarily within a single dependent project.

In contrast, the Awaitility library stands out, with only 28\% of its API methods (15 out of 53) falling into the low-usage category.  This suggests that a larger proportion of Awaitility's methods are used across multiple dependent projects, indicating broader adoption and more frequent usage of its features than other libraries.
On the other hand, a small percentage of API methods across all libraries—about 10\%  are categorized as highly used—meaning they are found in \rev{five or more} dependent projects. These methods are crucial to the library's ecosystem, and their changes could potentially affect many dependent projects.

Hence, we observe some diversity in distribution patterns across libraries. Still, we believe \rev{it}
helps developers identify which methods have a greater impact on their ecosystem and allows for more informed decision-making when considering breaking changes.  By understanding which API methods are critical to the ecosystem, library developers can better assess the potential risks and impacts of modifying or deprecating these methods.

\begin{quotebox}
    \textbf{Findings} Although library developers provide a wide variety of API methods (mean 4027), only 16\% are used by their 50 dependent projects. 
    Of the 16\% used API methods, only 7\% are used by more than \rev{9} of the 50 dependent projects.
   
\end{quotebox}

\begin{figure*}
    \centering
    \includegraphics[width=.95\textwidth]{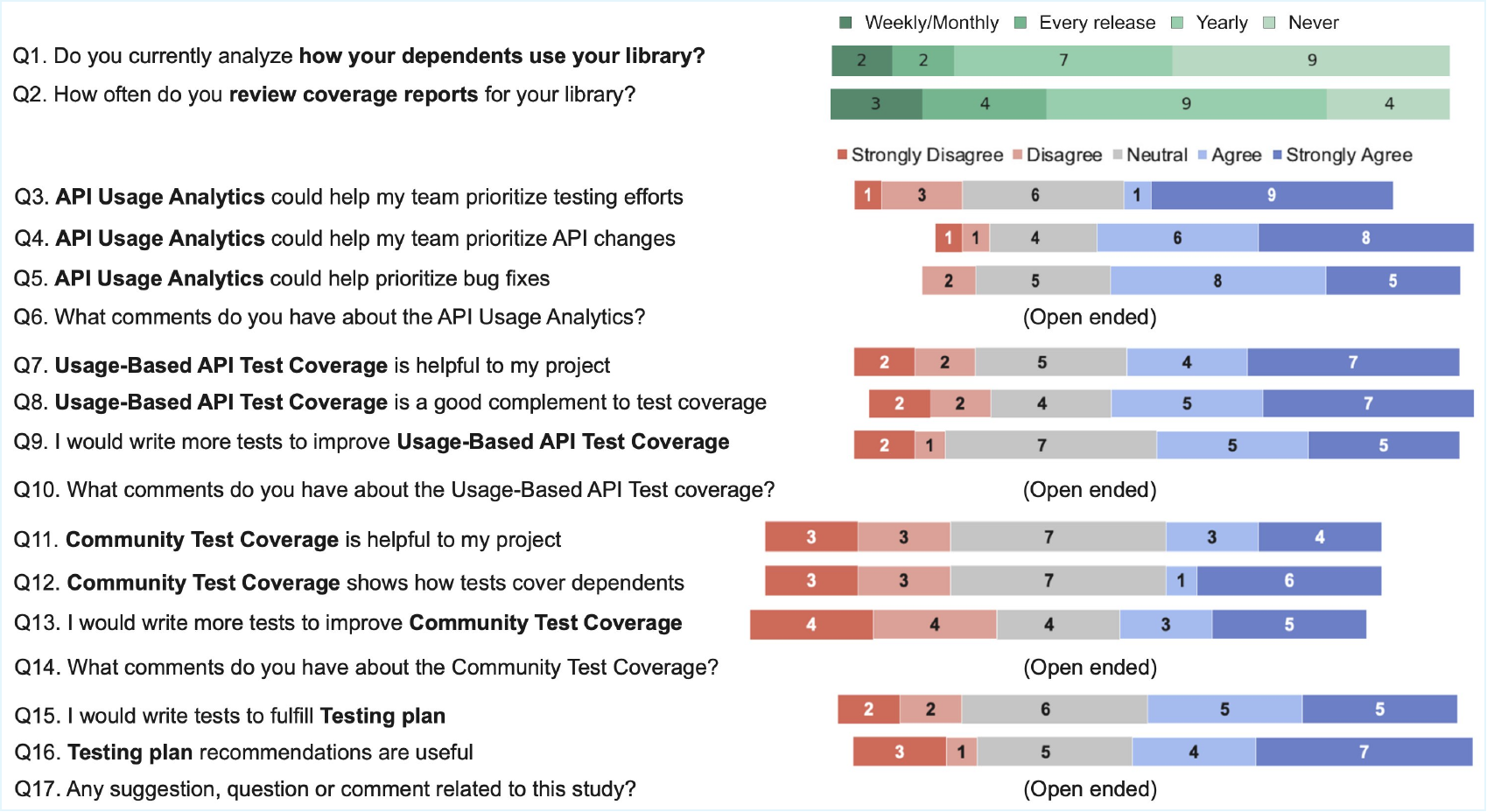}
    \vspace{-.2cm}
    \caption{\rev{Survey questions and results from the survey of 20 OSS contributors and maintainers.}}
    \label{fig:SurveyResults}
\end{figure*}

\subsection{Is API Usage Analytics useful for maintainers?}
\label{sub:api-analytics-usefulness}

To understand participants' feedback on API usage analytics, we first asked participants if they currently analyze how dependents use their library APIs (\textbf{Q1}). 
We find that only 4 (2\%) respondents stated to analyze this data at least once a month or every release. Most participants 7 (35\%) reported to analyzing dependents sporadically (once a year) or not at all.  
Then, we provided participants with two scenarios: (1) the top 10 most used API methods and their usage frequency across projects, and (2) the percentage of API methods used by 50 dependents relative to the total provided by the libraries, to help them understand the context and answer our questions. 

\rev{Figure \ref{fig:SurveyResults} presents the survey results for the quantitative questions and Table~\ref{tab:tikz-thematic-summary} showcases the themes extracted from the optional open-ended responses.}
Overall, we find that most respondents agree on the usefulness of using API usage analytics in the maintenance and evolution of their library project. Results show 50\% (10/20) agreement on that API Usage analytics could help prioritize testing efforts (\textbf{Q3}). 
A larger agreement, 70\% (14/20), was found when analysing if API Usage Analytics can help prioritize API changes (\textbf{Q4}).  
We find that 65\% (13/20) of respondents agree or strongly agree that having API Usage analytics could help prioritize bug fixes (\textbf{Q5}). \rev{This result is further reflected in the open-ended responses (\textbf{Q6}). 
As Table \ref{tab:tikz-thematic-summary} shows, four responses mention that API Usage Analytics can support API evolution: 
\textit{"Having the analytics of the most used APIs will help my team prioritize which APIs to update/change."}}
Three responses also mentioned that it can help prevent breaking changes.
Another respondent mentioned that knowing the most used APIs is important \textit{(...) mostly to find out what you cannot update, as too many people rely on it}.

However, 2\% (4/20) respondents disagreed, questioning the prioritization of testing efforts and bug fixes \rev{ (\textbf{Q3, Q4}). 
\textit{As a contributor, I would typically work first on what is relevant to me and secondly on the issues reported. Time left for studying how people use my library to add extra tests based on that would be difficult to find.}
}

\begin{quotebox}
    \textbf{Findings} Our survey showed that most participants agreed on the usefulness of API-Usage Analytics, particularly when prioritizing API changes (14/20), and bug-fixes (13/20). 
\end{quotebox}

\section{Usage-Based API Test Coverage}

\subsection{How often are used API methods covered by tests?} 
This section compares testing coverage and the API usage analytics extracted from the dependent projects. 
We extract the usage-based API test coverage using the methodology described in Section~\ref{sub:usage-based-api-test-coverage}.
Table \ref{tab:coverage} presents a comparison of the library project's line coverage and our proposed usage-based coverage across selected libraries.

Some libraries show discrepancies between the project statement coverage and the Usage-Based API Test Coverage. 
This may indicate that not all API methods used by dependent projects are covered by tests within the library’s own suite. 
For instance, Guava exhibits high statement coverage (94\%), yet the usage-based coverage is notably lower (65\%). Similarly, Javassist also shows a line coverage of 72\%, but only 50\% usage-based coverage. In Guava, 361 out of 1046 API method dependents lack test coverage, while in Javassist, 171 out of 281 remain untested. These gaps highlight potential blind spots where frequently used API methods are left unverified within the library’s own test suite, increasing the risk of uncaught API method regressions.

Conversely, some libraries exhibit a matched coverage between the project statement and usage-based API coverage. 
Commons-codec, AssertJ, and JUnit exhibit high usage-based coverage (92\%, 77\%, and 90\%, respectively), indicating that many of their widely used APIs are exercised in their test suites.

\begin{quotebox}
    {\textbf{Findings:} On average, 73.8\% of API methods used by dependents have partial or full test coverage. However, this still leaves gaps where some API methods used by dependents are not explicitly tested in the test suite.} 
\end{quotebox}

\subsection{Is Usage-Based API Test Coverage useful?}

For this part of the survey evaluation, \rev{we first asked how often maintainers review coverage reports for their projects (\textbf{Q2}). We find that 7 (35\%) review it weekly/monthly or every release. 13 (65\%) participants said that they do it once a year or never. We then} 
explained the proposed metric with either 1) the results of our analysis of their software project (specialized survey) or 2) a hypothetical scenario. 
Maintainers were then asked to answer Q7-Q10 based on their assessment of the metric's usefulness to their library projects. 

Figure \ref{fig:SurveyResults} presents the result of our survey.
Overall, maintainers showed a consistent agreement with the usefulness of the usage-Based API test coverage metric as a complementary coverage metric and a drive for more tests. 
Most participants, 55\% (11/20), agreed that the proposed coverage metric is helpful to their project (\textbf{Q7}), and 60\% (12/20) see the metric as a good complementary to conventional coverage metrics (\textbf{Q8}).
Half of the participants, 10/20 (50\%), were also receptive to the idea of writing more tests to improve the metric (\textbf{Q9}).
\rev{For the open-ended question (\textbf{Q10}), responses from table \ref{tab:tikz-thematic-summary} also highlights benefit of usage-based test coverage as 2 participants said "it helps cut tested but unused API features". However,}
A participant suggested variations of this metric, considering a weighted metric based on API usages stating that: \textit{Maybe the metric can be enhanced with some weights based on the number of dependents using each API method. An API used by more dependents should have more weight in the metric.}
\rev{Some respondents raised issues with chasing a type of coverage metric, stating that: 
\textit{Good coverage does not speak much about the quality of tests.}}
Another participant mentioned that equating partially covered APIs and fully covered may negatively impact the quality of the metric: \textit{The fact that the coverage can be partial or full, not sure if this is reliable. Better differentiate between fully and partially covered APIs.}

\begin{table}
\caption{Themes from the survey open-ended responses}
    \vspace{-.2cm}
    \label{tab:tikz-thematic-summary}
    \centering
    \newcolumntype{L}{>{\RaggedRight\arraybackslash}X}

\begin{tabularx}{\linewidth}{Lr}
    \toprule
    \rev{\textbf{API Usage Analytics}} & \rev{\textbf{\#}} \\
    \midrule

    \rev{\textit{(Benefits)} Supports API evolution by guiding developers to work on the most used API features}  & \rev{4} \\
    \rev{\textit{(Benefits)} Helps prevent breaking changes} & \rev{3} \\
    \rev{\textit{(Limitation)} Only works for popular libraries} & \rev{1} \\
    \midrule
    \midrule
    \rev{\textbf{Usage-Based Test Coverage}} & \rev{\textbf{\#}} \\
    \midrule
    \rev{\textit{(Benefit)} Helps cut tested but unused API features} & \rev{2} \\
    \rev{\textit{(Suggestion)} Includes metric weighted on the API usage} & \rev{1} \\
    \rev{\textit{(Limitation)} Coverage is not indicative of testing quality} & \rev{1} \\
    \rev{\textit{(Limitation)} JaCoCo is useful but lacks fine-grained analysis of execution paths} & \rev{1} \\     
    \rev{\textit{(Suggestion)} Differentiate partial and full test coverage} & \rev{1} \\
    \midrule
    \midrule
    \rev{\textbf{Community Test Coverage (CTC) and Testing Plan}} & \rev{\textbf{\#}} \\
    \midrule

    \rev{\textit{(Benefit)} Testing plans are useful and can drive actionability of coverage metrics} & \rev{4} \\
    \rev{\textit{(Limitation)} CTC is less applicable or useful than traditional test coverage} & \rev{2} \\
    \rev{\textit{(Limitation)} CTC is seen as outside of developers' control} & \rev{2} \\
    \rev{\textit{(Suggestion)} Suggestion of a time-based metric} & \rev{1} \\
    
\bottomrule     
\end{tabularx}

\end{table}

\begin{quotebox}
    \textbf{Findings} Most participants agree on the usefulness of Usage-Based API Test Coverage, as it can complement other coverage metrics (12/20) and drive the creation of new tests (10/20). 
\end{quotebox}

\begin{table}
\centering
\caption{Comparison of Project Statement Coverage and the Usage-Based API Test Coverage for the selected libraries.}
\vspace{-.2cm}
\begin{tabular}{l|r|r}
\toprule
\textbf{Library} & \makecell{\textbf{Project} \\ \textbf{Coverage (\%)}} & \makecell{\textbf{Usage-Based API} \\ \textbf{Test Coverage}} \\

\midrule
AssertJ & 92 & 1694/2210 (77\%) \\
Awaitility & 75 & 31/53 (77\%) \\
Commons Codec & 97 & 44/49 (92\%) \\
Guava & 94 & 685/1046 (65\%) \\
HttpComponents Client & 63 & 938/1308 (73\%) \\
Jackson-Databind & 70 & 154/216 (71\%) \\
Javassist & 72 & 110/281 (50\%) \\
Jsoup & 93 & 128/179 (72\%) \\
JUnit & 85 & 93/103 (90\%) \\
Logback & 74 & 366/509 (72\%) \\
\midrule
\textbf{Mean} & \textbf{81.5} & \textbf{424.8/595.4 (73.8\%)} \\
\bottomrule
\end{tabular}

\label{tab:coverage}
\vspace{-.2cm}
\end{table}

\section{Community Test Coverage}

\subsection{How often are dependent projects using only APIs fully covered by tests?}

\begin{table}
    \caption{Community test coverage (CTC) of the studied Java library projects. The testing plan showcases the impact of including tests to cover up to 10 highly-used API methods (``Tested APIs'' column) in the CTC (New CTC).}
    \centering
    \label{tab:community-test-coverage}
    \begin{tabular}{lr|rr}
    \toprule
    
    &  & \multicolumn{2}{c}{\textbf{Testing Plan}}  \\
    
    \textbf{Library} & \textbf{CTC} & \textbf{Tested APIs} & \textbf{New CTC} \\
    \midrule
    
    AssertJ & 30\% & 10 & 52\% \\
    Awaitility & 4\% & 10 & 94\% \\
    Commons Codec & 96\% & 2 & 100\% \\
    Guava & 4\% & 10 & 7\% \\
    HttpComponents Client & 7\% & 10 & 33\% \\
    Jackson-databind & 31\% & 10 & 56\% \\
    Javassist & 6\% & 10 & 24\% \\
    Jsoup & 76\% & 9 & 100\% \\
    JUnit & 54\% & 10 & 79\% \\
    Logback & 13\% & 10 & 40\% \\
    \midrule
    \textbf{Mean} & \textbf{32.1\%} & -- & \textbf{58.5\%} \\
    
    \bottomrule
\end{tabular}

    \vspace{-.2cm}
\end{table}

This section reports on the community test coverage of the studied library projects, using the method described in Section~\ref{sub:community-test-coverage}.
In summary, this metric captures how many of the 50 dependent projects rely only on API methods fully covered by the library test suite.

Table \ref{tab:community-test-coverage} presents the Community Test Coverage (CTC) values for the ten libraries.
On average, only 32\% of the dependent projects rely solely on fully covered API methods.
There is a high variability of the CTC across the studied library projects. 
Libraries like Commons-codec and JSoup have high CTC of 96\% and 76\%, meaning that most of their dependents rely on API methods explicitly exercised during the library test suite. 
On the other hand, libraries like Guava (4\%), Awaitility (4\%), and Javassist (6\%) exhibit a lower CTC, indicating that most of their dependents use one or more API methods that are not fully covered by the library’s tests.

\textbf{Proposing a testing plan:} 
To help libraries expand their test suite to cover more of the used API methods, we explored the impact of fully covering up to 10 API methods in the community test coverage. 
We ranked 10 API methods, previously not covered by the library test suite, based on their usage by the dependent ecosystem.
We then simulated the impact of the CTC (New CTC) when including these API methods in the test suite. 

Table \ref{tab:community-test-coverage} also showcases the result of this experiment, within the Testing Plan subcolumns. 
The findings show that targeted testing efforts in covering the dependent community can be quite effective in increasing the Community Test Coverage. 
Awaitility sees the most drastic improvement, with its CTC jumping from 4\% to 94\% by adding tests to just 10 API methods. 
This means that, by just including tests for 10 API methods, the library could virtually cover the APIs used by 50 dependent projects with at least one single test. 
HttpComponents Client, Javassist, and Logback also show significant improvements, increasing their CTC by 26–33 percentage points by including tests for up to 10 API methods. 
Other library projects, such as JSoup and Commons-Codec, can achieve full Community Test Coverage of their 50 dependent projects by adding tests to 2 and 9 API methods, respectively.

\begin{quotebox}
    \textbf{Findings:} On average, only 32\% of the dependent projects use only API methods fully covered by library test suites. Including up to 10 new highly-used API methods in the test suites can increase the CTC to 58\%.    
\end{quotebox}

\newcommand{\plusicon}{%
  \tikz[baseline=-0.5ex]
    \draw[green!50!black, line width=0.8pt] 
    (0,-0.1ex) -- (0,1ex) 
    (-0.5ex,0.45ex) -- (0.5ex,0.45ex);
}

\newcommand{\minusicon}{%
  \tikz[baseline=-0.5ex]
    \draw[red!70!black, line width=0.8pt] 
    (-0.5ex,0) -- (0.5ex,0);
}
\newcommand{\circleicon}{\tikz[baseline=-0.5ex]\draw[gray!70!black, line width=0.7pt] (0,0) circle (0.6ex);}

\newcommand{\neutralicon}{%
  \tikz[baseline=-0.5ex]
    \draw[gray!70!black, line width=0.7pt]
      (0,-0.6ex) -- (0,0.6ex)
      (-0.45ex,-0.45ex) -- (0.45ex,0.45ex)
      (-0.45ex,0.45ex) -- (0.45ex,-0.45ex);
}

\subsection{Is Community Test Coverage useful for maintainers?}

To evaluate its usefulness, we explain the idea behind the Community Test Coverage metric, including an example from their own library project (specialized survey) or a hypothetical scenario (general survey). 
Additionally, we proposed the idea behind the Testing Plan recommendation. 
In the specialized survey, we showed the results of the ranking of API method to improve Community Test Coverage.
In the general survey, we only present a hypothetical scenario to help them understand the concept.

Figure~\ref{fig:SurveyResults} presents the likert-scale results on community test coverage (CTC) metric, from Q11-Q14 and testing plan from Q15-Q17. 
Overall, the community test coverage was received with mixed assessment. 
In \textbf{Q11}, we observe equal distribution in the agreement spectrum among participants (7 agreements, 7 neutral and 6 disagreements) when evaluating if knowing CTC can be helpful for their library project.
Similar agreement levels were obtained when assessing if CTC is a reasonable metric to represent how library tests cover their dependents (\textbf{Q12}), and whether maintainers would write more tests to increase the CTC (\textbf{Q13}). \rev {By analysing the open-ended question (\textbf{Q14}), we assessed benefits and some of their core concerns/limitations. Table \ref{tab:tikz-thematic-summary} presents the thematic coding for CTC and testing plan. }

A core concern raised by participants is that community-based coverage relies on how dependents use the library APIs, which are seen by maintainers as outside of their control. 
\textit{"A quality metric outside my control doesn't sound like something I would want to make architectural decisions on."}
Furthermore, we noticed some echoed concerns with the efficacy of coverage metrics in general, similarly of what we received in the Usage-Based API Test Coverage: 
\textit{It doesn't cover how they use my library. There is a huge amount of variation in ordering, inputs and other setup that majorly impacts the results. Only knowing that I have written a few tests (with useful assertions) that hit a method doesn't help me much.}

On the other hand, participants have higher agreement on the usefulness of our derived \textbf{Testing Plan}.
When presented with the impact of increasing the test coverage to 10 highly-used API methods, half the participants 11/20 (55\%) agree that the testing plan is useful (\textbf{Q16}).
Additionally, we find that 10/20 participants agreed that, if recommended such a plan, they would write tests to cover the highly-used API methods (\textbf{Q15}). 
\rev{Comments for open-ended question (\textbf{Q17}) were positive on the actionability of such a test plan recommendation. As seen in Table~\ref{tab:tikz-thematic-summary}, 4 participants specifically stated that testing plan can help drive actionability of coverage metrics. A participant stated: \textit{"I like that the code coverage metric will become more actionable if a good testing plan is generated out of it."}}
Another participant mentioned the clear goal of a testing plan as a positive factor: \textit{"It is an easy goal to explain to new contributors, it has an immediate effect, and it has a definition of done you are working towards. I can see that being quite motivating to a lot of people."}.

\begin{quotebox}
    \textbf{Findings} Maintainers have mixed opinions on the usefulness of the Community Test Coverage metric.
    Participants frequently agreed on the usefulness of the derived testing plan (11/20), which can help them prioritize testing new API methods (10/20).
\end{quotebox}

\section{Discussions \& Implications}
\label{discussion}

\subsection{Towards a community-based tool.}

\textbf{Challenges \& limitations:} 
A notorious limitation of our approach is that it relies on an ecosystem of dependent projects to provide feedback to library maintainers. 
Our approach will likely apply better to well-established libraries with large dependent ecosystems, while newer or niche library projects may not benefit from an ecosystem-level analysis. 
There is also an important concern about which dependent projects to include in the community pool of dependent ecosystems. 
Analysing low-quality dependent projects may also yield unreliable results and confuse maintainers.

We use static analysis tools to analyze the source code of dependent projects hosted on GitHub.
This approach allowed us to include both library and application. However, we faced certain known limitations in using code analysis to map all API usages such as type-solving and particularities of the Java language (e.g., object polymorphism, method overload) \cite{li2017understandinganalyzingjavareflection, 7985689}.
This may pose challenges for performing large-scale analysis needed to provide a reliable analytics tool for open-source Java library projects. We envision that byte-code analysis of the dependent project's Jar files could yield a more precise and unambiguous analysis. This analysis could be performed in the ecosystem of Java libraries through the Maven registry. At the cost of skipping Java applications, the byte-code analysis may be a more reliable approach in mining community-based analytics from large sets of dependent projects.

\rev{\textbf{Potential Applicability of CTC}}
\rev{Community Test Coverage (CTC) is developed to help library maintainers assess and improve the adequacy of their test suites from the perspective of real-world usage. Unlike traditional coverage metrics that focus on unit tests, CTC shifts the focus to external impact. Achieving 100\% internal coverage is often impractical, but CTC offers a practical alternative by measuring how many dependent projects rely only on fully tested API methods. Our idea is to provide maintainers with a community-centric perspective on test coverage. When widely used API methods are fully covered, regressions are more likely to be caught early, protecting dependents from unexpected breakages and reducing the maintenance burden.}

\rev{Like established metrics from tools like SonarQube, CTC can be integrated into CI workflows to produce actionable reports and targeted test recommendations, as demonstrated in our study. However, its effectiveness is highest for mature libraries with stable APIs and a sizable, observable dependent base. Nonetheless, it can be used for any library as long as API usage can be observed, making the CTC a valuable and practical complement to traditional coverage metrics.} 

\textbf{Future ideas based on maintainers' feedback: }
Our survey showed that maintainers are interested in community-based analytics; however, several respondents desired more advanced features to derive better actionability.
One suggestion was to present a timeline of API usage, rather than using the latest snapshot of dependent projects. This would allow maintainers to analyse the history of API usages from the dependent ecosystem and spot trends of API migration and adoption of the new API features.
Respondents also encouraged moving beyond frequency-based analysis of individual API methods, proposing instead that usage be grouped by library features.
Overall, the feedback highlights strong interest in community-based tools and points to several promising research opportunities to better support open-source library maintainers.

\subsection{Coverage as a test quality metric}

Two of our community-based metrics are related to evaluating test effectiveness through the lens of test coverage. 
It is widely understood that test coverage is not strongly correlated with test effectiveness~\cite{inozemtseva2014coverage,testcoverageondefectcoverage}, and can steer developers to pursue irrelevant metrics~\cite{ayoup:2022:achievementunlocked}.
However, while high test coverage is not a good indicator of good test quality, the lack of coverage remains a useful indicator of potential gaps in the test suite~\cite{brandttowards}. 

Recent studies have raised the question of identifying critical parts of the software that tests should cover, attempting to guide developers in prioritizing testing efforts~\cite{aniche2022effective, ivankovic2024productive}. 
In this study, we approach a similar problem for library maintainers: increasing awareness of the testing gaps for used API methods. 
By combining test coverage with API Usage, we can help increase the awareness of testing gaps that could affect a large set of their dependent ecosystem. 
This analysis can highlight \rev{those blind spots to} maintainers, where testing could be improved to make the library more resilient.

\section{Threats to Validity}
\label{sec:ttv}

\rev{\textbf{Generalization of the results.} 
The sample size of ten open-source projects limits the generalizability of the findings related to API usage and test coverage reported in our study. 
It is unclear whether the patterns we observe with the analytics (API Usage and test coverage) will hold on a larger sample size, and it is unlikely to generalize to the entire Java ecosystem. 
However, the goal of our study is to demonstrate the applicability and usefulness of community-based analytics for open-source library maintainers across a variety of domains (e.g., testing, logging). 
As such, we believe the survey results of 20 experienced OSS maintainers from widely used Java library projects pose valuable insights for our research community in exploring this avenue further.
}

\textbf{Sample size of dependent projects:}
While the libraries we studied have tens of thousands of dependents, we filtered for projects using the latest major version to ensure relevance to the current API. This choice, while narrowing the dependent pool, allows us to focus feedback on the most up-to-date library usage. Despite the smaller sample size, the selected 50 projects offer a meaningful representation of how developers interact with popular libraries. While a broader set might reveal additional API methods, we find this sample sufficient to capture the most frequently used APIs for our analysis.

\noindent
\textbf{Internal threat to validity:} 
We encountered challenges configuring JavaParser to handle Java files across multiple versions, limiting our analysis to Java 8–14 and 17. Files from unsupported versions could not be processed due to incomplete support in JavaParser. Due to this, some \rev{API} methods could not be resolved during analysis and were therefore skipped. While this reduced the number of methods identified, it yielded a conservative estimate, avoiding false positives. Resolved methods were correctly identified, lending confidence to our results. 
Although JaCoCo is a mature tool, it has known limitations \cite{soto2023coverage}, such as incomplete coverage of fields, methods that only throw exceptions, and compiler-generated code. However, these limitations are unlikely to affect our study, as we focus on public API methods with executable logic.

\textbf{Uncertainty in API methods mapping}: We faced challenges when mapping methods extracted from dependent projects (API Usage) to their respective test coverage report.
In 14\% of the extracted API methods (Partial Matching with Ambiguity), we found multiple potential matches when mapping API methods to the test coverage report. 
In these cases, we decided to map the API method to the highest test coverage, applying an optimistic (\rev{upper}-bound) coverage report in the Usage-Based API Test Coverage and the Community Test Coverage. 
While we believe our results hold, they should be interpreted as a \rev{upper}-bound value. It is likely that the real coverage of API methods could be lower than reported.

\section{Conclusion}

In this paper, we introduce community-based metrics and analytics to help open-source maintainers make informed decisions about their libraries. 
Our approach includes API usage analysis, where we examine how dependent projects utilize library API methods and compare this usage with the library's test suite coverage. 
Our empirical analysis of ten popular Java libraries showed the potential for a community-based analytics tool to help maintainers: 
understand the gap between the features libraries provide and what their ecosystem uses, 
use this insight to prioritize maintenance efforts.

We surveyed open-source maintainers and contributors to understand the usefulness of community-based analytics in guiding their efforts in maintaining and evolving their library projects.
Survey responses from 20 participants indicate that maintainers see value in API usage analytics and the metric on Usage-Based API Test Coverage. 
Opinions on the Community Test Coverage metric are mixed, but maintainers found our testing plan recommendations helpful for improving overall project test coverage.
We believe this study is a step towards bridging the gap between OSS maintainers and their dependent ecosystem.

\definecolor{darkgreen}{rgb}{0.0, 0.5, 0.0}
\definecolor{darkred}{rgb}{0.5, 0.0, 0.0}

\bibliographystyle{ACM-Reference-Format}
\bibliography{bibliography}
\appendix

\twocolumn

\end{document}